# Citing is earlier than Cited?

Xian Li

Medical Information Center, Zhejiang University School of Medicine, Hangzhou 310058, China

**Abstract:** Generally, it is common that cited papers are earlier than citing papers. But we found three different cases, with more undiscovered. In this letter, we attempted to explain the reasons. However, negative time lag between citing and cited papers may mislead us when we study the characteristics of science.

The statement "*If I have seen further, it is by standing on the shoulders of giants.*" said by Newton is widely used to describe that science development replies on the previous researches. In fact, scientists need to cite related researches to support their points and references list is helpful to understand knowledge flow from cited papers to citing papers. Obviously, the publication time of cited papers must be earlier than citing papers. However, I found some different cases when searching for information in Web of Science, the world's most trusted publisher-independent global citation database[i], via the library of Zhejiang University. For example, "*a novel multigene family may encode odorant receptors - a molecular-basis for odor recognition*" with "10.1016/0092-8674(91)90418-X" published in 1991 was cited by "*biotechnology of beta-adrenergic receptors*" published in 1990. Similarly, the second paper published in 1979 with "10.1073/pnas.76.9.4275" was cited by "*pharmacology and endogenous roles of prostaglandin endoperoxides, thromboxane-a2, and prostacyclin*" in 1978. And the last paper I found published in 1955 with "10.1042/bj0590438" was cited by "*\*le systeme hexose-phosphatasique .4. specificite de la glucose-6-phosphatase*" in 1954. Fig 1 shows yearly citation of cited papers.

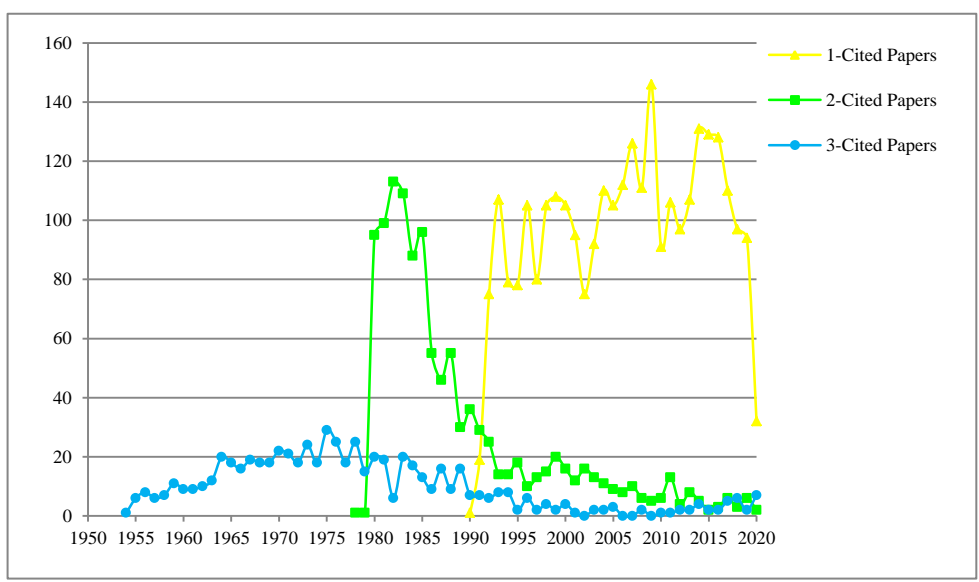

Fig 1 Yearly citation of cited papers

Therefore, an interesting question was put forward-"Why would citing papers be earlier than cited papers?" As far as I'm concerned, this could happen in a few ways. 1) Citing and cited authors know each other or they met at a conference and they talked about the cited papers. 2) Cited authors as reviewers suggested citing authors to cite their upcoming relevant articles. 3) Cited papers are shown online before official publication, and citing authors could have spotted and cited the official publication date rather than the online date.

In this letter, although we found three papers with negative time lag between citing and cited papers (TLCC), I believe that there are still more papers with the same phenomenon. As we know, TLCC as an indicator represents the speed of knowledge diffusion (Nakamura et al, 2011). However, negative TLCC might mislead us when we investigate the knowledge diffusion pattern. Besides, citing authors could have encountered citation bribery (Ross, 2019) so that they have to cite unpublished article. Therefore, not only editors but also authors are supposed to pay more attention the correct use of citation.

---

[i] More details can be found at: https://clarivate.com/webofsciencegroup/solutions/web-of-science